\documentclass[showpacs,twocolumn,prb,eqsecnum,amsmath,amssymb]{revtex4}
\usepackage{graphicx}
\usepackage{dcolumn}
\usepackage{bm}

\begin{document}
\title{Mott metal-insulator transition in the Hubbard model}
\author{Fusayoshi J. Ohkawa}
\affiliation{Department of Physics, Faculty of Science, 
Hokkaido University, Sapporo 060-0810, Japan}
\email{fohkawa@phys.sci.hokudai.ac.jp}
\date{\today}
\begin{abstract} 
The Hubbard model in the strong-coupling regime is mainly studied by Kondo-lattice 
theory or $1/d$ expansion theory, with $d$ being the spatial dimensionality.
In two dimensions and higher, the ground state within the Hilbert subspace
with no order parameter is a normal Fermi liquid except for $n=1$ and
$U/W=+\infty$, with $n$ being the electron density per unit cell, $U$ the
on-site repulsion, and $W$ the bandwidth; the cooperation between the Kondo effect, 
which favors a local singlet on each unit cell, and a resonating-valence-bond effect, 
which favors a local singlet on each pair of nearest-neighbor unit cells,
stabilizes the Fermi liquid, whose ground state is a singlet as a whole,
in the strong-coupling regime. In the whole Hilbert space with no restriction, 
the normal Fermi liquid is unstable at least against a magnetic or
superconducting state. This analysis confirms an early Fermi-liquid theory
of high-temperature superconductivity, F. J. Ohkawa, Jpn. J. Appl. Phys.
{\bf 26}, L652 (1987). 
The ground state for $n=1$ and $U/W=+\infty$ is a
Mott insulator. Actual metal-insulator transitions cannot be explained
within the Hubbard model.  In order to explain them, the electron-phonon
interaction, multi-band or multi-orbital effects, and effects of  disorder
should be considered beyond the Hubbard model. 
The crossover between
local-moment magnetism and itinerant-electron magnetism corresponds to that
between a localized spin and a normal Fermi liquid in the Kondo
effect and it is simply a Mott metal-insulator crossover. 
\end{abstract}
 %
\maketitle

\section{Introduction}
\label{SeIntroduction}

The Mott metal-insulator (M-I) transition is an interesting and important
issue in solid-state physics,\cite{mott} and a lot of effort has been made
towards clarifying it. \cite{tokura} However, its theoretical treatment is
still controversial. One of the most contentious issues is whether or not
the transition can be explained within the Hubbard model.

In the Hubbard approximation, \cite{Hubbard1,Hubbard2} provided that the
on-site repulsion $U$ is large enough such that $U\agt W$ or $W/U\alt1$,
with $W$ being the bandwidth, a band splits into two subbands or the Hubbard
gap opens between the upper Hubbard band (UHB) and the lower Hubbard band
(LHB). In the Gutzwiller approximation,
\cite{Gutzwiller1,Gutzwiller2,Gutzwiller3} a narrow band of
quasi-particles appears around the chemical potential; the band and
quasi-particles are called the Gutzwiller band and quasi-particles in this
paper. One may speculate that the density of states in fact has a
three-peak structure, with the Gutzwiller band between UHB and LHB. Both
of the approximations are single-site approximations (SSA). Another SSA
theory confirms this speculation, \cite{OhkawaSlave} showing that the
Gutzwiller band appears at the top of LHB when the electron density per
unit cell $n$ is less than one, i.e., $n<1$. According to Kondo-lattice
theory, \cite{Mapping-1,Mapping-2,Mapping-3} the three-peak structure
corresponds to the Kondo peak between two subpeaks in the Anderson model,
which is an effective Hamiltonian for the Kondo effect. An insulating
state appears provided that not only the Hubbard gap opens but also the
Fermi surface of the Gutzwiller quasi-particles vanishes.

Provided that $n=1$ and $W/U=+0$, an electron is localized at a unit cell 
and it behaves as a free localized spin, so that the ground state is
infinitely degenerate and is a typical Mott insulator. This fact implies
that the ground state is also a Mott insulator in the vicinity of $n=1$
and $W/U=+0$, as is also implied by experiment. However, there is an
argument that contradicts this implication: For example, assume that a
nonzero but infinitesimally small density of electrons are removed from
the Mott insulator or {\it holes} are doped into the Mott insulator. It is
reasonable that the holes are itinerant at $T=0$~K provided that no gap
opens in the Gutzwiller band and no disorder exists. 

\begin{figure}
\centerline{\includegraphics[width=5.5cm]{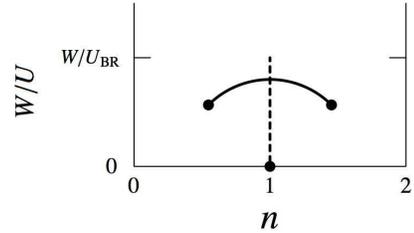}}
\caption[1]{
Schematic phase diagram of the ground state within the Hilbert subspace
with no order parameter in two dimensions and higher. The ground state is
a metal except for $n=1$ and $W/U=+0$. The arc indicates a
possible but unlikely first-order transition line between metallic states, 
as is discussed in Sec.~\ref{SecMMtransition}. 
Dots indicate critical points. 
The dashed line indicates an insulator line implied by Brinkman and Rice's theory,
\cite{brinkman} but it cannot survive when a resonating-valence -bond effect is
considered, as is examined in Sec.~\ref{SecBroadening}}
\label{fig_phase}
\end{figure}


In the Gutzwiller approximation, when $W/U=+0$ the effective mass of the
quasi-particles diverges as
$n\rightarrow 1 \pm 0$. When $n\ne 1$, in fact, electrons are itinerant
even for $W/U=+0$. According to Brinkman and Rice's theory,
\cite{brinkman} which is also under the Gutzwiller approximation, when
$n=1$ the effective mass diverges as 
$U\rightarrow U_{\rm BR}-0$, with $U_{\rm BR} \simeq W$.
It is implied that, within the Hilbert subspace with no order
parameter, the ground state is an insulator for $n=1$ and $0\le W/U \le
W/U_{\rm BR}$, i.e., on the dashed line in the phase diagram shown in
Fig.~\ref{fig_phase}. 
The divergence of the effective mass occurs continuously, so that the M-I
transition is of second order. It is unconventional that no order
parameter appears in this second-order transition and no discontinuity
seems to occur across the dashed line, which implies that the critical
$U_{\rm BR}$ is infinite beyond the Gutzwiller approximation such that
$W/U_{\rm BR}\rightarrow +0$.

One of the purposes of this paper is to show that no Mott M-I transition
is possible at any finite $U$. Since actual M-I transitions cannot be
explained within the Hubbard model, another purpose is to examine relevant
effects for the transitions beyond the Hubbard model. The other purpose
is to examine two issues related with the Mott M-I transition: the
crossover between local-moment magnetism and itinerant-electron magnetism
and high-temperature (high-$T_c$) superconductivity in cuprate oxides.
\cite{ bednortz} This paper is organized as follows: The ground states
within SSA and beyond SSA are studied in Secs.~\ref{SecSSA} and 
\ref{SecBeyondSSA}, respectively. Relevant effects in actual M-I
transitions are considered in Sec.~\ref{SecBeyondHub}. The magnetism
crossover is considered in Sec.~\ref{SecMagnetismCrossover}. High-$T_c$
superconductivity is considered in Sec.~\ref{SecSuperconductivity}.
Discussion is given in Sec.~\ref{SecDiscussion}. Conclusion is given in
Sec.~\ref{SecConclusion}. A proof of an inequality, which plays a critical
role in this paper, is given in Appendix~\ref{SecProof}. When cuprate
oxide superconductors approach the Mott M-I transition or crossover, the
specific heat coefficient $\gamma$ is suppressed 
\cite{loram,momono} and tunneling spectra are asymmetric, \cite
{asymmetry1} both of which are unconventional. A phenomenological
analysis on these issues is given in Appendix~\ref{SecAppendixRho}.

\section{Fermi liquid within SSA}\label{SecSSA}
\subsection{Fermi-surface condition}

The Hubbard model is defined by 
\begin{equation}\label{EqHubbard} {\cal H} = \epsilon_a
\sum_{i\sigma}n_{i\sigma} - \sum_{i\ne j}\sum_{\sigma} t_{ij}
a_{i\sigma}^\dag a_{j\sigma} + U \sum_{i} n_{i\uparrow} n_{i\downarrow},
\end{equation} 
with $n_{i\sigma}= a_{i\sigma}^\dag a_{i\sigma}$. The notations are
conventional here.
The dispersion relation of electrons is given by 
\begin{equation}\label{EqBareDispersion}
E({\bf k}) = \epsilon_a -
\frac1{N} \sum_{i\ne j} t_{ij} \exp\left[i{\bf k}\cdot
\left({\bf R}_i-{\bf R}_j\right)\right] ,
\end{equation}
with $N$ being the number of unit cells and
${\bf R}_i$ the position of the $i$th lattice site.
The density of states 
as a function of the electron energy $\varepsilon$ is defined by
\begin{equation}\label{EqRho0}
\rho_0(\varepsilon) = \frac1{N} \sum_{\bf k}
\delta[\varepsilon - E({\bf k})] ,
\end{equation}
and, for convenience, the density of states 
as a function of the electron density $n$ is defined by 
\begin{subequations}\label{EqRhoN0}
\begin{equation}
\bar{\rho}_0(n) = \frac1{N} \sum_{\bf k}
\delta[\mu_0 (n)- E({\bf k})] ,
\end{equation}
with $\mu_0(n)$ defined by
\begin{equation}
n = 2 \int_{-\infty}^{\mu_0(n)} d\varepsilon \rho_0(\varepsilon).
\end{equation}
\end{subequations}
An effective bandwidth of $E({\bf k})$ or $\rho_0(\varepsilon)$ is denoted 
by $W$ in this paper. It is assumed that the Fermi surface (FS) is present
for $U=0$ or $\bar{\rho}_0(n) >0 $ for any $0<n<2$.

As is discussed in Introduction, the Kondo effect has relevance to
electron correlations in the Hubbard model. The $s$-$d$ model is one of
the simplest effective Hamiltonians for the Kondo effect. According to
Yosida's perturbation theory \cite{yosida} and Wilson's
renormalization-group theory, \cite{wilsonKG} provided that FS of
conduction electrons is present, the ground state of the $s$-$d$ model is
a singlet or a normal Fermi liquid (FL) but is exceptionally a doublet for
$J_{s\mbox{-}d}=0$, with $J_{s\mbox{-}d}$ the $s$-$d$ exchange interaction.
The FL is stabilized by the Kondo effect or the quenching of magnetic
moments by local quantum spin fluctuations.

The $s$-$d$ model is derived from the Anderson model, which is defined by
\begin{eqnarray}\label{EqAnderson}
{\cal H}_{A}&=&
\sum_{{\bf k}\sigma} E_c({\bf k}) 
c_{{\bf k}\sigma}^\dag c_{{\bf k}\sigma}
+ \epsilon_d \sum_{\sigma}n_{d\sigma}
+ \tilde{U} n_{d\uparrow} n_{d\downarrow}
\nonumber \\ && 
+ \frac1{\sqrt{N_{A}}} \sum_{{\bf k}\sigma} \left[
V({\bf k})c_{{\bf k}\sigma}^\dag d_\sigma
+ (\mbox{h.c.}) \right] ,
\end{eqnarray}
 with $n_{d\sigma}=d_{\sigma}^\dag d_{\sigma}$ and $N_{A}$ the number of
unit cells. The notations are also conventional here.
The hybridization energy is defined by
\begin{equation}\label{EqL}
L_\sigma(i\varepsilon_n) =
\frac1{N_{A}} \sum_{\bf k}
\frac{|V({\bf k})|^2 }{
i\varepsilon_n+\tilde{\mu} - E_c({\bf k}) },
\end{equation}
with $\tilde{\mu}$ being the chemical potential. It follows that
\begin{equation}
\mbox{Im}\left[L_\sigma(\varepsilon+i0) \right]=
- \frac{\pi}{N_A} \sum_{\bf k} |V({\bf k})|^2 
\delta\left[\varepsilon+\tilde{\mu }- E_c({\bf k})\right], 
\end{equation}
A necessary and sufficient condition for the presence of FS is simply given by
\begin{subequations}\label{EqFL}
\begin{equation}\label{EqFL1}
\mbox{Im}\left[L_\sigma(+i0) \right] < 0.
\end{equation} 
When $\mbox{Im}\left[L_\sigma(\varepsilon+i0) \right]$ is discontinuous at
$\varepsilon=0$,
\begin{equation}\label{EqFL2}
\lim_{\varepsilon\rightarrow \pm 0} 
\mbox{Im}\left[L_\sigma(\varepsilon+i0) 
\right] < 0, 
\end{equation}
\end{subequations}
is more relevant than Eq.~(\ref{EqFL1}). The condition (\ref{EqFL1}) or
(\ref{EqFL2}) is called the FS condition in this paper.
 According to the
result on the $s$-$d$ model, \cite{yosida,wilsonKG} provided that the FS condition
is satisfied, the ground state of the Anderson model is
a singlet or a normal FL but is exceptionally a doublet for the just half
filling and infinite $\tilde{U}$.

When there is no order parameter,
the Green function of the Hubbard model is given by
\begin{equation}\label{EqGreen}
G_{\sigma}(i\varepsilon_n, {\bf k}) =
\frac1{i\varepsilon_n + \mu - E({\bf k}) - 
\Sigma_{\sigma}(i\varepsilon_n, {\bf k}) },
\end{equation}
with $\mu$ the chemical potential of the Hubbard model
and $\Sigma_{\sigma}(i\varepsilon_n, {\bf k})$ 
the single-particle self-energy. 
The self-energy 
is divided into single-site and multi-site self-energies:
\begin{equation}
\Sigma_{\sigma}(i\varepsilon_n, {\bf k})
= \tilde{\Sigma}_{\sigma}(i\varepsilon_n)
+ \Delta \Sigma_{\sigma}(i\varepsilon_n, {\bf k}).
\end{equation}
Provided that the on-site interaction and the single-site electron lines 
are the same in the Feynman diagrams of the Hubbard and Anderson models,
the single-site $\tilde{\Sigma}_{\sigma}(i\varepsilon_n)$
is given by that of the Anderson model.
The condition for the on-site interaction is simply given by
$\tilde{U}=U$.
The single-site Green function of the Hubbard model is given by
\begin{equation}\label{EqR}
R_{\sigma}(i\varepsilon_n) =
\frac1{N} \sum_{\bf k} G_{\sigma}(i\varepsilon_n{\bf k}),
\end{equation}
and that of the Anderson model is given by
\begin{equation}
\tilde{G}_{\sigma}(i\varepsilon_n) =
\frac1{
\displaystyle i\varepsilon_n + \tilde{\mu}
- \epsilon_d - \tilde{\Sigma}_{\sigma}(i\varepsilon_n)
-L_\sigma(i\varepsilon_n) },
\end{equation}
with $L_\sigma(i\varepsilon_n) $ defined by Eq.~(\ref{EqL}).
The condition for the electron lines is simply given by
\begin{equation}
R_{\sigma}(i\varepsilon_n) = \tilde{G}_\sigma (i\varepsilon_n).
\end{equation}
In fact, a set of $\tilde{U}=U$, 
$\tilde{\mu}-\epsilon_d = \mu - \epsilon_a $, and
 \begin{equation}\label{EqMappingCondition}
 L_\sigma(\varepsilon + i0)=\varepsilon + \mu -\epsilon_a
- \tilde{\Sigma}_\sigma(\varepsilon + i0) 
-\frac1{R_\sigma(\varepsilon + i0)},
\end{equation}
is a mapping condition to the Anderson model. A problem of calculating the
single-site $\tilde{\Sigma}_\sigma(\varepsilon + i0)$ is reduced to a
problem of determining and solving self-consistently the Anderson model.
\cite{Mapping-1,Mapping-2,Mapping-3}


When the multi-site
$\Delta \Sigma_{\sigma}(i\varepsilon_n, {\bf k})$ is ignored in the
mapping condition~(\ref{EqMappingCondition}), the approximation is the 
best SSA because it considers all the single-site terms. The SSA is
rigorous for infinite dimensions within the Hilbert subspace with no order
parameter. \cite{Metzner} It can also be formulated as the dynamical
mean-field theory \cite{georges,RevMod,kotliar,PhyToday} (DMFT) and the
dynamical coherent potential approximation. \cite{dpca}

\subsection{Adiabatic continuation}
The multi-site $\Delta \Sigma_{\sigma}(i\varepsilon_n, {\bf k})$ is 
ignored in the following part of this section. Consider a Lorentzian model
or the Hubbard model with a Lorentzian density of states:
\begin{equation}
\rho_0(\varepsilon)=
\frac1{\pi} \frac{\Delta}{ (\varepsilon-\epsilon_a)^2 + \Delta^2 } ,
\end{equation}
with $\Delta= W/\pi$. Then, Eq.~(\ref{EqR}) is simply given by
\begin{equation}\label{EqR-Lorentz}
R_\sigma(\varepsilon+i0) =
\frac1{\varepsilon +\mu - \epsilon_a 
+ i \Delta - \tilde{\Sigma}_\sigma(\varepsilon+i0)} .
\end{equation}
In principle, the mapping condition~(\ref{EqMappingCondition}) should be
treated in an iterative manner to determine the Anderson model to be
solved. However, no iteration is needed for this model because 
Eq.~(\ref{EqMappingCondition}) gives \cite{georges}
\begin{equation}
L_\sigma(\varepsilon + i0) = - i\Delta,
\end{equation}
even when any input $\tilde{\Sigma}_{\sigma}(\varepsilon+i0)$ is used
in the right side of Eq.~(\ref{EqMappingCondition}).
The SSA is simply reduced to solving the Anderson model. 
Since the FS condition (\ref{EqFL}) is satisfied for the Anderson model,
the ground state of the Hubbard model is a normal FL
except for $n=1$ and $W/U=+0$.

One may argue that an M-I transition at finite $U$ is only possible when
$\rho_0(\varepsilon)$ has finite band-tails. 
In order to examine a non-Lorentzian model of $\rho_0(\varepsilon)$,
which may have finite or infinite band-tails, the following model is 
first examined:
\begin{equation}
\rho_{\delta} (\varepsilon) = - \frac1{\pi} \mbox{Im}
\int d\varepsilon^\prime \frac{\rho_0(\varepsilon^\prime)}
{\varepsilon-\varepsilon^\prime + i\delta\Delta},
\end{equation}
with $\delta > 0$. In this non--Lorentzian model,
\begin{eqnarray}\label{EqR-delta}
R_\sigma(\varepsilon+i0) &=& 
%
\int d\varepsilon^{\prime} \rho_0(\varepsilon^{\prime} ) 
\nonumber \\ && \hspace*{-1cm} \times
\frac1{\varepsilon +\mu - \epsilon_a - \varepsilon^{\prime} 
+ i \delta \Delta 
- \tilde{\Sigma}_\sigma(\varepsilon+i0)} ,\qquad 
\end{eqnarray}
instead of Eq.~(\ref{EqR-Lorentz}).
As is proved in Appendix~\ref{SecProof}, 
\begin{equation}\label{EqImp}
\mbox{Im}L_\sigma(\varepsilon+i0) \le - \delta \Delta ,
\end{equation}
for any input $\tilde{\Sigma}_{\sigma}(\varepsilon+i0)$. For example,
one may argue a possible scenario for a Mott insulator with a nonzero gap 
across the chemical potential is that the self-energy develops a pole at
$\varepsilon=0$ such that
\begin{equation}
\tilde{\Sigma}_{\sigma}(\varepsilon+i0) =
c_p\frac{\Delta^2}{\varepsilon+i0} + \cdots ,
\end{equation}
with $c_p$ a numerical constant. Even if this type of the self-energy is
tried as an input of the iterative process in order to search a
self-consistent non-normal FL solution, $L_\sigma(\varepsilon+i0)$ given by
the mapping condition (\ref{EqMappingCondition}) satisfies
Eq.~(\ref{EqImp}). Since the FS condition (\ref{EqFL}) is satisfied
without fail in each iterative process to determine the Anderson model,
no non-normal  FL  solution can be obtained in the SSA theory 
or the ground state of an eventual self-consistent SSA solution should be a
normal FL. Provided that $\delta>0$, no M-I transition occurs at finite
$U$. The ground state for $\rho_{\delta}(\varepsilon)$ with $\delta>0$
is a Mott insulator only at $n=1$ and $W/U=+0$.



An SSA solution for $\rho_{0}(\varepsilon)$ is obtained by the adiabatic
continuation \cite{AndersonText} of $\delta \rightarrow+0$. Provided that
\begin{equation}
\lim_{\varepsilon\rightarrow \pm 0}\lim_{\delta\rightarrow +0} 
\mbox{Im}L_\sigma(\varepsilon+i0)< 0 ,
\end{equation}
the ground state of the SSA solution  is definitely a singlet or a normal FL.
On the other hand, provided that
\begin{equation}\label{EqViolateFS}
\lim_{\varepsilon\rightarrow \pm 0}\lim_{\delta\rightarrow +0} 
\mbox{Im}L_\sigma(\varepsilon+i0) = - 0 ,
\end{equation}
the ground state may be degenerate. The nature of the possible degeneracy is
examined in Sec.~\ref{SecDegeneracy}. 

\subsection{Fermi-liquid relation}
\label{SecFL-Relation}
First, consider the Anderson model self-consistently determined in the
absence of any external field, and apply infinitesimally small Zeeman
energy $g\mu_BH$ and chemical potential shift $\Delta\mu$ to the Anderson
model; Weiss mean fields induced by the external fields are not included
in this treatment. It is obvious that, provided that $\delta>0$, the
adiabatic continuation \cite{AndersonText} as a function of $U$ also holds. 
Therefore, the self-energy of the Anderson model for $\delta=+0$ is expanded 
in such a way that
\begin{eqnarray}\label{EqExpandSelf}
\tilde{\Sigma}_\sigma(\varepsilon + i0) &=&
\tilde{\Sigma}_0(0)
+ \left(1 - \tilde{\phi}_\gamma \right)\varepsilon 
+ \left(1 - \tilde{\phi}_s \right) \frac1{2}\sigma g \mu_BH 
\nonumber \\ && 
+ \left(1 - \tilde{\phi}_c \right) \Delta\mu
 + O\left(\varepsilon^2\right), \qquad 
\end{eqnarray}
at $T=0$~K, with $\tilde{\Sigma}_0(0)$, $\tilde{\phi}_\gamma$,
$\tilde{\phi}_s$, and $\tilde{\phi}_c$ all being real. According to the
Fermi-liquid relation,\cite{yosida-yamada} the specific heat coefficient is
given by
\begin{equation}\label{EqFLR-G}
\gamma = \frac{2}{3} \pi^2 k_B^2 \tilde{\phi}_\gamma \rho^*(0).
\end{equation}
Here, $\rho^*(0)$ or $\rho^*(\varepsilon)$ is the density of states 
defined by
\begin{equation}\label{EqRhoSSA}
\rho^*(\varepsilon) = -\frac1{\pi}\mbox{Im} 
\tilde{G}_\sigma(\varepsilon+i0) 
= -\frac1{\pi}\mbox{Im} R_\sigma(\varepsilon+i0) .
\end{equation}
Static spin and charge susceptibilities are given by
\begin{equation}\label{EqFLR-S}
\tilde{\chi}_s(0) = 
2\tilde{\phi}_s \rho^*(0) ,
\end{equation}
and 
\begin{equation}\label{EqFLR-C}
\tilde{\chi}_c(0) = 
2\tilde{\phi}_c \rho^*(0) ,
\end{equation}
respectively. The conventional factor $(1/4)g^2\mu_B^2$ is not included
in $\tilde{\chi}_s(0)$. It also follows that \cite{yosida-yamada}
\begin{equation}\label{EqPhiSum}
2\tilde{\phi}_\gamma =\tilde{\phi}_s+\tilde{\phi}_c.
\end{equation}
Since the on-site $U$ is repulsive, 
local charge fluctuations are suppressed, so that
\begin{equation}\label{EqSupC}
0< \tilde{\phi}_c/\tilde{\phi}_\gamma < 1.
\end{equation}
Then, it follows that
\begin{equation}\label{EqEnhanceS}
1< \tilde{\phi}_s/\tilde{\phi}_\gamma < 2. 
\end{equation}
It is likely that $\tilde{\phi}_c/\tilde{\phi}_\gamma \ll1$ and
$\tilde{\phi}_s/\tilde{\phi}_\gamma \simeq 2$ for $n\simeq 1$ and
$U/W\agt 1$. The Kondo temperature, which is the energy scale of local
quantum spin fluctuations, is defined by
\begin{equation}\label{EqDefTK}
k_BT_K = \left[1/\tilde{\chi}_s(0) \right]_{T=0~{\rm K}} .
\end{equation}


The self-energy of the Hubbard model in the absence of any external field
is simply given by $\tilde{\Sigma}_\sigma(\varepsilon + i0)$ with 
$g\mu_BH=0$ and $\Delta\mu=0$. The density of states for the Hubbard model
is the same as that for the Anderson model model, as is shown in
Eq.~(\ref{EqRhoSSA}). According to the Fermi-liquid relation,
\cite{Luttinger1,Luttinger2} the specific heat coefficient of the Hubbard
model is also given by Eq.~(\ref{EqFLR-G}). Local spin and charge
susceptibilities of the Hubbard model are given by Eqs.~(\ref{EqFLR-S})
and (\ref{EqFLR-C}). The energy scale of local quantum spin fluctuations
in the Hubbard model is also the Kondo temperature defined by
Eq.~(\ref{EqDefTK}).

According to the FS sum rule, \cite{Luttinger1,Luttinger2} 
the electron density $n$ is given by
\begin{equation}\label{EqFS-SumRule}
n = \frac1{N} \sum_{{\bf k}\sigma}
\theta\Bigl( [ \mu - \epsilon_a - E({\bf k}) - \tilde{\Sigma}_0(0)
]/W \Bigr) ,
\end{equation}
with $\theta(x)$ being the step function defined by
\begin{equation}
\theta(x) = \left\{\begin{array}{cc}
0, & x <0 \\
1, & x>0
\end{array}\right. .
\end{equation}
According to Eqs.~(\ref{EqRhoN0}) and (\ref{EqFS-SumRule}),
it follows that
\begin{equation}\label{EqFS-SumRule1}
\mu - \epsilon_a -\tilde{\Sigma}_0(0) = \mu_0(n) .
\end{equation}
According to Eq.~(\ref{EqFS-SumRule}) or (\ref{EqFS-SumRule1}), 
provided that $n$ is kept constant, $\mu -\tilde{\Sigma}_0(0)$, 
$R_\sigma(+i0) $, $\rho^*(0)$, and $L_{\sigma}(+i0)$ 
do not depend on $U$. It should be noted that
\begin{subequations}\label{EqNonExotic}
\begin{eqnarray}\label{EqRhoSSA2}
\rho^*(0) &=&
\frac1{N} \sum_{{\bf k}\sigma}
\delta\left[ \mu - \epsilon_a - E({\bf k}) - \tilde{\Sigma}_0(0) 
\right] 
\nonumber \\ &=& \bar{\rho}_0(n) >0, 
\end{eqnarray}
and
\begin{equation}\label{EqImL-D}
\mbox{Im}L_\sigma(+i0) = 
- \frac{\pi\rho^*(0)}{\left[\mbox{Re}R_\sigma(+i0)\right]^2 + 
\left[\pi\rho^*(0) \right]^2} <0.
\end{equation}
\end{subequations}

The dispersion relation and an effective bandwidth of the quasi-particles 
are defined, respectively, by 
\begin{equation}\label{EqDisperSSA}
\xi_0({\bf k})= \frac1{\tilde{\phi}_\gamma}
\left[\epsilon_a + E({\bf k}) +\tilde{\Sigma}_0(0) - \mu \right],
\end{equation}
and 
\begin{equation}
W^*=W/\tilde{\phi}_\gamma.
\end{equation}
The Green function (\ref{EqGreen}) is approximately 
divided into the so called coherent and incoherent terms:
\begin{equation}\label{EqGreen0}
G_\sigma(i\varepsilon_n, {\bf k}) =
\frac1{\tilde{\phi}_\gamma}
\frac1{i\varepsilon_n - \xi_0({\bf k})} + [\mbox{incoherent term}].
\end{equation}
Here, the first term is the coherent term, which describes the quasi-particle band,
and the incoherent term describes LHB and UHB.

\subsection{Possible degeneracy}
\label{SecDegeneracy}
Equation~(\ref{EqImL-D}) shows that the FS condition (\ref{EqFL1}) is
satisfied by the SSA solution for $\delta= +0$, as is expected. When both
of $\rho_0(\varepsilon)$ and 
$\tilde{\Sigma}_\sigma(\varepsilon + i0)$ are continuous and finite,
$L_\sigma(\varepsilon + i0)$ is continuous so that the FS condition
(\ref{EqFL2}) is also satisfied. In such a case, the ground state is
never degenerate and is simply a normal FL.
On the other hand, when $\rho_0(\varepsilon)$ or 
$\tilde{\Sigma}_\sigma(\varepsilon + i0)$ is discontinuous or divergent,
$L_\sigma(\varepsilon + i0)$ can be discontinuous so that it is possible
that the FS condition (\ref{EqFL2}) is not satisfied or
Eq.~(\ref{EqViolateFS}) is satisfied, Eq.~(\ref{EqImL-D}) notwithstanding.

When $\rho_0(\varepsilon)$ is discontinuous or divergent at
$\varepsilon=\mu_0(n)$, $\mbox{Re}R_\sigma(\varepsilon+i0)$ or 
$\rho^*(\varepsilon)$ is divergent at $\varepsilon=0$. Then,
Eq.~(\ref{EqViolateFS}) is satisfied so that the ground state may be
degenerate. When $\rho_0(\varepsilon)$ is divergent at 
$\varepsilon=\mu_0(n)$, the ground state is degenerate even for $U=0$.

Since $\tilde{\phi}_\gamma$ is finite in Eq.~(\ref{EqExpandSelf}) provided
that $\delta>0$, only the possible scenario for the discontinuity or
divergence of $\tilde{\Sigma}_\sigma(\varepsilon + i0)$ at $\varepsilon=0$
is that $\tilde{\phi}_\gamma\rightarrow +\infty$ as $\delta\rightarrow +0$.
In such a case, the real part of $\tilde{\Sigma}_\sigma(\varepsilon + i0)$
is at least discontinuous at $\varepsilon=0$; it may be finite or
divergent as $\varepsilon \rightarrow \pm 0$. When the real part is
discontinuous, the imaginary part exhibits logarithmic divergences as
$\varepsilon \rightarrow \pm 0$ according to the Kramers-Kronig relation. 
Provided that $\tilde{\phi}_\gamma\rightarrow +\infty$ as $\delta\rightarrow +0$,
it follows that
\begin{subequations}\label{EqExotic}
\begin{equation}
\lim_{\varepsilon\rightarrow\pm0} 
\lim_{\delta\rightarrow+0}
\rho^*(\varepsilon) = + 0 , 
\end{equation}
and
\begin{equation}
\lim_{\varepsilon\rightarrow\pm0} 
\lim_{\delta\rightarrow+0}
\mbox{Im}L_\sigma(\varepsilon+i0) = - 0.
\end{equation}
\end{subequations}
It should be noted that Eq.~(\ref{EqNonExotic}), which is for
$\varepsilon=0$, still holds.
In the exceptional case of $n=1$ and $W/U=+0$,
\begin{subequations}\label{EqExoticC}
\begin{equation}
\rho^*(\varepsilon) = 0 , 
\end{equation}
and
\begin{equation}
\mbox{Im}L_\sigma(\varepsilon+i0) =0,
\end{equation}
\end{subequations}
for any finite $\varepsilon$, and for any $\delta\ge +0$.

There are three possible scenarios for the phase diagram: When the
divergence of $\tilde{\phi}_\gamma$ occurs as $\delta\rightarrow +0$ at a
point on the $\delta=0$ plane, the point is a critical point. When it
occurs as $\delta\rightarrow +0$ at any point on a line, the line is a
critical line. When it occurs as $\delta\rightarrow +0$ at any point on a
plane, the plane is a critical plane. The transition is of second order in
any scenario.

It is unlikely that there is an isolated critical point of $n\ne1$ or
$W/U>+0$. When the scenario of a critical point is the case, the critical
point should be the point of $n=1$ and $W/U=+0$. The critical point is
exotic because there is discontinuity in $\rho^*(\varepsilon)$ as a
function of $\varepsilon$, $n$, and $W/U$ at the critical point, as is
shown in Eqs.~(\ref{EqNonExotic}), (\ref{EqExotic}), and (\ref{EqExoticC}).
The critical line and plane on the $\delta=0$ plane are more exotic than
the critical point is. They should include the point of 
$n=1$ and $W/U=+0$ as a critical point within themselves. Then, there is
discontinuity in $\rho^*(\varepsilon)$ as a function of $\varepsilon$,
$n$, and $W/U$ at the critical point even within the critical line and
plane. 

According to Eqs.~(\ref{EqFLR-G}), (\ref{EqEnhanceS}), and (\ref{EqDefTK}),
$\gamma\rightarrow +\infty$ ~mJ/mol~K$^2$ and $T_K \rightarrow +0$~K as 
$\tilde{\phi}_\gamma\rightarrow +\infty$, which simply means that 
low-energy or zero-energy states are accumulated or the ground state is
degenerate. The divergence of the local spin susceptibility
$\tilde{\chi}_s(0)$ is also one of the consequences of the degeneracy of 
the ground state. At the critical point of $n=1$ and $W/U=+0$, an electron
behaves as a free localized spin so that $\tilde{\chi}_s(0) = 1/k_BT$,
which diverges as $T\rightarrow 0$~K. A similar divergent behavior is
expected on the critical line or plane.

In a conventional second-order phase transition, not only an order
parameter and infinite degeneracy of the ground state but also rigidity
appear so that a ground-state configuration is rigidly realized among
infinitely degenerate ones; the Nambu-Goldstone mode appears and the
entropy is zero at $T=0$~K. Only an external field conjugate to the order
parameter can lift the degeneracy of the ground state. The transition
discussed here, which is also of second order, is quite different from the
conventional one. No order parameter or no rigidity appears so that the
Nambu-Goldstone mode does not appear and the entropy is nonzero at
$T=0$~K, i.e., the third law of thermodynamics does not hold. An
infinitesimally small perturbation such as $\delta=+0$ can easily lift
the degeneracy or the degenerate ground state is not rigid against an
infinitesimally small perturbation. These unconventional features are 
totally obvious or trivial for the critical point of $n=1$ and $W/U=+0$.

When the ground state is degenerate, rigorously speaking, the FL is not a
normal FL. However, since Eq.~(\ref{EqRhoSSA2}) is satisfied even for
$\delta=+0$ and no order parameter or no rigidity appears, an SSA solution
with $T_K=+0$~K can be regarded as a normal FL with a vanishing effective
Fermi energy. {\em In fact, if $\tilde{\phi}_\gamma$ is extremely large
but is still finite for an extremely small but nonzero $\delta$, an SSA
solution for such a small $\delta$ is a normal FL with an extremely small
but nonzero Fermi energy.}

In the Gutzwiller approximation, \cite{Gutzwiller1,Gutzwiller2,Gutzwiller3}
when $W/U=+0$ it follows that $\tilde{\phi}_\gamma \propto 1/|1-n|$, which
implies that the scenario of a critical plane is unlikely. Then, Brinkman
and Rice's theory \cite{brinkman} implies the existence of the critical
line of $n=1$ and $0\le W/U \le W/U_{\rm BR}$, as is discussed in
Introduction; it is obvious that no discontinuity can occur across the
critical line. The degenerate ground state on the critical line is not
rigid, as is discussed above. It is therefore speculated that, provided
that $\rho_0(\varepsilon)$ is continuous and finite at
$\varepsilon=\mu_0(n)$, the critical line cannot survive in an SSA beyond
the Gutzwiller approximation; it cannot survive beyond SSA, as is examined
in Sec.~\ref{SecBroadening}.

\subsection{Possible first-order metal-metal transition}
\label{SecMMtransition}
It is assumed so far that a self-consistent SSA solution is unique. If it
is not unique, a first-order transition between metallic states is
possible. However, the adiabatic continuation still holds, for example,
along a route around one of the critical points at the ends of the
first-order transition line. Consider two metallic states that are on
different sides of the line but are infinitesimally close to each other. 
Since $n$'s are the same in two metallic states, the FS sum rule,
$\rho^*(0)$, and $L_\sigma(+i0)$ are all the same in the two metallic
states. It is difficult to imagine that, for example,
$\tilde{\phi}_\gamma$ shows a jump across the line. The occurrence of such
a first-order transition is unlikely. The transition never occurs in the
Lorentzian model because the mapping is unique. The transition line is
shown on a schematic phase diagram in Fig.~\ref{fig_phase}, although it
is unlikely.

\section{Ground state beyond SSA}
\label{SecBeyondSSA}
\subsection{Kondo-lattice or $1/d$ expansion theory}
\label{SecKondo1/D}
The irreducible spin polarization function is also divided into
single-site and multi-site functions:
\begin{equation}
\pi_s(i\omega_l,{\bf q}) =
\tilde{\pi}_s(i\omega_l) +\Delta\pi_s(i\omega_l,{\bf q}) .
\end{equation}
The single-site $\tilde{\pi}_s(i\omega_l)$ is given by that of the
Anderson model. The spin susceptibilities of the Anderson and Hubbard
models are given, respectively, by
\begin{equation}
\tilde{\chi}_s(i\omega_l) =
\frac{2\tilde{\pi}_s(i\omega_l) }{ 
1 - U \tilde{\pi}_s(i\omega_l) }, 
\end{equation}
and
\begin{equation}
\chi_s(i\omega_l,{\bf q}) =
\frac{2\pi_s(i\omega_l,{\bf q}) }{
1 - U \pi_s(i\omega_l,{\bf q}) }.
\end{equation}
A physical picture for Kondo lattices is that local spin fluctuations on
different sites interact with each other by an intersite exchange
interaction. In Kondo-lattice theory, according to this physical picture,
an intersite exchange interaction $I_s(i\omega_l,{\bf q})$ is defined by
\begin{equation}\label{EqKondoKai}
\chi_s(i\omega_l,{\bf q}) =
\frac{\tilde{\chi}_s(i\omega_l) }{
1 - \frac1{4} I_s(i\omega_l,{\bf q}) \tilde{\chi}_s(i\omega_l) }.
\end{equation}
Provided that $U/W\agt 1$, it follows that
\begin{equation}\label{EqExch-I}
I_s(i\omega_l,{\bf q}) = 2 U^2 \Delta\pi_s(i\omega_l,{\bf q})
\left[ 1 +O\left(\frac1{U\tilde{\chi}_s(i\omega_l) }\right) \right] ,
\end{equation}
where terms of $O[1/U\tilde{\chi}_s(i\omega_l)]$ can be ignored. The
strong coupling case of $U/W\agt 1$ is mainly studied in this section. 

The exchange interaction $I_s(i\omega_l,{\bf q}) $ is composed of three 
terms: \cite{three-exchange,itinerant-ferro}
\begin{equation}
I_s(i\omega_l,{\bf q}) =
J_s({\bf q}) + J_Q(i\omega_l,{\bf q})
- 4 \Lambda (i\omega_l,{\bf q}).
\end{equation}
The first term $J_s({\bf q})$ is the superexchange interaction. According
to field theory, it arises from the exchange of a pair excitation of
electrons between LHB and UHB. \cite{sup-exchange} When the widths of LHB
and UHB are vanishingly small, the strength of the superexchange
interaction between nearest neighbors is
$J = - 4|t|^2/U$,
with $t$ the transfer integral between nearest neighbors. Since the
widths of LHB and UHB are nonzero, 
$|J|$ becomes substantially smaller than $4|t|^2/U$, for example, about a
half of $4|t|^2/U$ in a realistic condition.
\cite{exchange-reduction} 

The second term $J_Q(i\omega_l,{\bf q})$ is an exchange interaction
arising from the exchange of a pair excitation of the quasi-particles.
According to the Ward relation, \cite{ward} the static component of the
single-site irreducible three-point vertex function in spin channels is
given by 
\begin{eqnarray}\label{Eq3pointVertex}
\tilde{\lambda}_s &=& \tilde{\phi}_s[1 -U \tilde{\pi}_s(0)]
\nonumber \\ &= &
\frac{2 \tilde{\phi}_s}{U \tilde{\chi}_s(0)} 
\left[ 1 +O\left(\frac1{U\tilde{\chi}_s(0) }\right) \right], 
\end{eqnarray}
where terms of $O[1/U\tilde{\chi}_s(0)]$ can also be ignored. When only
the coherent part of the Green function is considered and this
$\tilde{\lambda}_s$ is approximately used for low-energy dynamical
processes, $J_Q(i\omega_l,{\bf q})$ is given by
\begin{equation}\label{EqJQ}
J_Q(i\omega_l,{\bf q}) = P(i\omega_l,{\bf q}) 
- \frac1{N} \sum_{\bf q} P(i\omega_l,{\bf q}) ,
\end{equation}
with
\begin{eqnarray}
P(i\omega_l,{\bf q}) &=&
\frac{4}{\tilde{\chi}_s^2(0)}
\left(\frac{\tilde{\phi}_s}{\tilde{\phi}_\gamma}\right)^2
\nonumber \\ && \hspace*{-0.5cm} \times 
\frac1{N}\sum_{{\bf k}\sigma}
\frac{f[\xi_0({\bf k}) ]-f[\xi_0({\bf k} + {\bf q})]}
{i \omega_l - \xi_0({\bf k} + {\bf q}) + \xi_0({\bf k}) }, \qquad 
\end{eqnarray}
with
$f(\varepsilon) = 1/[e^{\varepsilon/k_BT} +1]$.
 In Eq.~(\ref{EqJQ}), the single-site term is subtracted because it is
considered in SSA. The strength of this exchange interaction is 
proportional to $1/\tilde{\chi}_s(0) = k_BT_K$, which is proportional to
the quasi-particle bandwidth.\cite{satoh1,satoh2} It is antiferromagnetic
when the nesting of FS is sharp or the chemical potential lies around the
center of the quasi-particle band. It is ferromagnetic when the chemical
potential lies around the top or bottom of the quasi-particle band. In
particular, it is strongly ferromagnetic when the density of states has a
sharp peak at one of the band edges where chemical potential lies, 
\cite{itinerant-ferro,satoh1,satoh2,miyai} as it has a sharp peak in many
itinerant-electron ferromagnets such as Fe, Ni, and so on.

The third term $- 4 \Lambda (i\omega_l,{\bf q})$ corresponds to the
mode-mode coupling term of spin fluctuations in the self-consistent
renormalization (SCR) theory, \cite{moriya} which is relevant for 
$U/W\alt 1$. 

When the three-point vertex function $\tilde{\lambda}_s$ given by
Eq.~(\ref{Eq3pointVertex}) is approximately used for low-energy dynamical
processes, the mutual interaction between the quasi-particles is given by
\begin{equation}\label{EqChi-J}
\frac1{4} (U \tilde{\lambda}_s)^2
[\chi_s(i\omega_l,{\bf q}) - \tilde{\chi}_s(i\omega_l)]=
\frac1{4} \tilde{\phi}_s^2 I_s^*(i\omega_l,{\bf q}), 
\end{equation}
with
\begin{equation}
I_s^*(i\omega_l,{\bf q})= \frac{ I_s(i\omega_l,{\bf q})}{
1 - \frac1{4} I_s(i\omega_l,{\bf q}) \tilde{\chi}_s(i\omega_l)}.
\end{equation}
In Eq.~(\ref{EqChi-J}), the single-site term is subtracted because it is
considered in SSA, and two $\tilde{\phi}_s$ appear as effective
three-point vertex functions. It should be noted that the mutual
interaction mediated by spin fluctuations is essentially the same as that
due to the exchange interaction $I_s(i\omega_l,{\bf q})$ or
$I_s^*(i\omega_l,{\bf q})$.

In Kondo-lattice theory, an {\em unperturbed} state is constructed in the
non-perturbative SSA theory and intersite effects are perturbatively
considered in terms of $I_s(i\omega_l,{\bf q})$ or $I_s^*(i\omega_l,{\bf
q})$. Kondo-lattice theory can also be formulated as $1/d$ expansion
theory,\cite{Mapping-2,Mapping-3} with $d$ the spatial dimensionality. 
What remain nonzero in the limit of $d\rightarrow +\infty$ are the
single-site self-energy $\tilde{\Sigma}_\sigma(i\varepsilon)$, the
single-site polarization function $\tilde{\chi}_s(i\omega_l)$, and the
magnetic exchange interactions, $J_s({\bf Q})$ and 
$J_Q(i\omega_l, {\bf Q})$, for particular ${\bf Q}$'s in the Brillouin
zone; both of $J_s({\bf q})$ and $J_Q(i\omega_l, {\bf q})$ vanish for
almost all ${\bf q}$'s. When the N\'{e}el temperature $T_N$ is nonzero,
magnetization ${\bf m}({\bf Q})$ appears at $T<T_N$. Therefore, 
$J_s({\bf Q}){\bf m}({\bf Q})$ and 
$J_Q(i\omega_l, {\bf Q}){\bf m}({\bf Q})$ can be nonzero even in the
limit of $d\rightarrow +\infty$, which are Weiss mean fields. All the
other terms such as $\Delta\Sigma_{\sigma}(i\varepsilon_n, {\bf k})$ and
$- 4 \Lambda (i\omega_l,{\bf q})$ vanish in the limit of 
$d\rightarrow +\infty$. \cite{com1/D}

\subsection{Stabilization of the normal Fermi liquid}
\label{SecBroadening}
The quasi-particles are renormalized by the intersite exchange interaction
$I_s^*(i\omega_l,{\bf q})$. One of the main terms of $I_s^*(i\omega_l,{\bf
q})$ is the superexchange interaction:
\begin{eqnarray}
I_s^*(i\omega_l,{\bf q}) &=& 
I_s(i\omega_l,{\bf q})
+ \frac{ \frac1{4} I_s^2(i\omega_l,{\bf q})\tilde{\chi}_s(i\omega_l)}{
1 - \frac1{4} I_s(i\omega_l,{\bf q}) \tilde{\chi}_s(i\omega_l)}
\nonumber \\ &=&
J_s({\bf q}) + J_Q(i\omega_l,{\bf q}) + \cdots .
\end{eqnarray}
There are two types of the renormalization linear in the superexchange
interaction. One is a Hartree-type term,\cite{mag-structure}
$\tilde{\phi}_s J_s({\bf Q}){\bm m}({\bf Q})$, which may cause magnetic
instability. In this subsection, it is not considered in order to restrict
the Hilbert space within the subspace with no order parameter; possible
instabilities are examined in Sec.~\ref{SecInstabilityFL}. The other is
a Fock-type term, which stabilizes the FL, as is examined below.

When only the coherent term of the Green function is considered, 
the Fock-type term is given by\cite{comFullSelfconsistet} 
\begin{equation}\label{EqSelf-1}
\Delta \Sigma_\sigma ({\bf k}) = 
\frac{3}{4} \frac{\tilde{\phi}_{\rm s}^2}{\tilde{\phi}_\gamma}
\frac{k_B T}{N} 
\sum_{\varepsilon_{n^\prime}{\bf k}^\prime}
J_s({\bf k}- {\bf k}^\prime) 
\frac{ e^{i \varepsilon_{n^\prime}0^+} }
{i\varepsilon_n ^\prime- \xi_0({\bf k}^\prime)} .
\end{equation}
Here, the factor 3 appears because of three spin channels and two
effective vertex functions $\tilde{\phi}_{\rm s}^2$ appear. When the
multi-site self-energy is considered in the mapping condition
(\ref{EqMappingCondition}), the single-site and multi-site terms depend on
each other. In principle, therefore, they should also be self-consistently
calculated with each other.
Once $\tilde{\phi}_\gamma$, $\tilde{\phi}_s$, and $\Delta \Sigma_\sigma
({\bf k})$ are self-consistently calculated, the dispersion relation of
the quasi-particles is given by
\begin{equation}\label{EqDisperBeyond}
\xi({\bf k})= \frac1{\tilde{\phi}_\gamma}
\left[\epsilon_a + E({\bf k}) +\tilde{\Sigma}_0(0) 
+ \Delta\Sigma_\sigma({\bf k}) - \mu \right],
\end{equation}
and the density of states at the chemical potential by
\begin{equation}\label{EqRhoBeyondSSA}
\rho^*(0) = \frac1{N}\sum_{\bf k} \delta 
\left[\epsilon_a + E({\bf k}) +\tilde{\Sigma}_0(0) 
+ \Delta\Sigma_\sigma({\bf k}) - \mu \right] .
\end{equation}
When this $\rho^*(0)$ is used instead of Eq.~(\ref{EqRhoSSA2}), 
the specific heat coefficient is given by Eq.~(\ref{EqFLR-G}) and 
the local spin susceptibility is given by Eq.~(\ref{EqFLR-S})

The renormalization (\ref{EqSelf-1}) depends on dimensionality $d$ and 
the lattice structure. When only the superexchange interaction $J$
between nearest neighbors is considered, for example, in a square-lattice
model, it follows that
\begin{equation}\label{EqDeltaSigma}
\frac1{\tilde{\phi}_\gamma}\Delta \Sigma_\sigma ({\bf k}) =
\frac{3}{4} \left(\frac{\tilde{\phi}_{\rm s}}{\tilde{\phi}_\gamma}\right)^2
J \Xi \left[\cos(k_xa) +\cos(k_ya)\right] , 
\end{equation}
with $a$ the lattice constant, and
\begin{equation}\label{EqXi}
\Xi =
\frac1{ N}\sum_{{\bf k}} \theta \left[-\frac{\xi_0({\bf k})}{W}\right]
\left[\cos(k_xa) +\cos(k_ya)\right] .
\end{equation}
Since $1<\tilde{\phi}_s/\tilde{\phi}_\gamma<2$, as is shown in
Eq.~(\ref{EqEnhanceS}), Eq.~(\ref{EqDeltaSigma}) remains nonzero even if
$\tilde{\phi}_\gamma$ is divergent. In general, when an effective
bandwidth of 
$\Delta\Sigma_\sigma({\bf k})/\tilde{\phi}_\gamma$ is denoted by 
$c_J |J|$, an effective bandwidth of $\xi({\bf k})$ is given by 
\begin{equation}\label{EqW*}
W^* = \frac{W}{\tilde{\phi}_\gamma} + c_{J} |J| ,
\end{equation}
with $c_J =O(1)$ being a numerical constant, which depends on $d$ 
and the lattice structure.

When this renormalization is considered, it follows that
\begin{equation}\label{EqRho9}
\rho^*(0) \simeq \frac1{\displaystyle W + \tilde{\phi}_\gamma c_{J} |J| },
\end{equation}
and
\begin{equation}
k_BT_K = 
\frac1{2\tilde{\phi}_s\rho^*(0)} 
\simeq \frac{\displaystyle W + \tilde{\phi}_\gamma c_{J} |J| }
{2\tilde{\phi}_s}.
\end{equation}
It should be noted that the Kondo temperature $T_K$ is nonzero even if
$\tilde{\phi}_s \rightarrow +\infty$ or
$\tilde{\phi}_\gamma \rightarrow +\infty$, provided that $|J|$ is nonzero. 
Since the vanishment of $T_K$ and the divergence of $\tilde{\phi}_\gamma$ 
occur together in any case provided that the ground state is degenerate,
the fact that $T_K$ can never be zero leads to a conclusion that the
divergence of $\tilde{\phi}_\gamma$ can never occur provided that
$\tilde{\phi}_\gamma$ is self-consistently calculated beyond SSA. The
degeneracy of the ground state never occurs except for $n\rightarrow1$ and
$W/U\rightarrow +0$. Even if the critical line or plane is present under
SSA, it can never survive beyond SSA. It is trivial that
the critical point $n=1$ and $W/U=+0$ survives.

It follows according to Eq.~(\ref{EqRho9}) that
\begin{equation}
\lim_{\tilde{\phi}_\gamma \rightarrow +\infty} \rho^*(0) = +0,
\end{equation}
for $|J|\ne 0$ or $W/U>0$. Excepting on the line of $W/U=+0$, there is no
discontinuity in $\rho^*(\varepsilon)$ as a function of $\varepsilon$,
$n$, and $W/U$. However, there is still a discontinuity at $n=1$ on the
line of $W/U=+0$. This discontinuity presumably vanishes when the
renormalization by the total $I_s^*(i\omega_l,{\bf q})$ is considered.
The critical point of $n=1$ and $W/U=+0$ is a conventional one beyond SSA. 

When the superexchange interaction between nearest neighbors is strong
enough but no antiferromagnetic order occurs, the quasi-formation of a
singlet on each pair of nearest-neighbor unit cells occurs or local
quantum spin fluctuations are developed on each pair of nearest-neighbor
unit cells. The Fock-type term considers effectively the quenching effect
of magnetic moments by the spin fluctuations, which stabilizes the normal
FL. In fact, the FL reached or constructed by the adiabatic continuation
under SSA, which is stabilized by the quenching of magnetic moments by
{\em single-site} local quantum spin fluctuations, is further stabilized 
by that by {\em nearest-neighbor} local quantum spin fluctuations. The
phase diagram of the ground state is shown in Fig.~\ref{fig_phase}, which
applies even to one dimension at least under the approximation where only
the Fock-type term is considered beyond SSA; the Fock-type term is never
divergent even in one dimension.

\subsection{Instability of the Fermi liquid}
 \label{SecInstabilityFL}
 An order parameter can appear in two dimensions and higher. The
instability of the normal FL can be examined when the response function
corresponding to the order parameter is perturbatively considered in terms
of $I_s(i\omega_l,{\bf q})$ or $I_s^*(i\omega_l,{\bf q})$.

Since the main term of $I_s(i\omega_l,{\bf q})$ is the superexchange
interaction, most possible order parameters are simply what can be derived
from the decoupling of
\begin{equation}
{\cal H}_J = - \frac1{2} J\sum_{\left<ij\right>}
\sum_{\nu}\sum_{\alpha\beta\gamma\delta}
\left(\mbox{$\frac1{2}$}\sigma_{\nu}^{\alpha\beta}\right)
\left(\mbox{$\frac1{2}$}\sigma_{\nu}^{\gamma\delta}\right)
a_{i\alpha}^\dagger a_{i\beta} a_{j\gamma}^\dagger a_{j\delta },
\end{equation}
with the summation $\left<ij\right>$
being over nearest-neighbor sites and
$\sigma_{\nu}^{\alpha\beta}$ ($\nu=x$, $y$, and $z$)
being the Pauli matrixes.
Three types of order parameters are possible in the mean-field
approximation. The first is a magnetic order parameter, which is given by 
$\sum_{\tau\tau^\prime} \sigma_{\nu}^{\tau\tau^\prime}
\bigl< a_{i\tau}^\dagger a_{i\tau^\prime }\bigr>$. The second is a
superconducting (SC) one, which is given by 
$\sum_{\tau\tau^\prime} \bigl< a_{i\tau}^\dagger a_{j\tau^\prime }^\dagger
\bigr> $ for nearest-neighbor $\left<ij\right>$. The third is a bond-order
(BO) one; charge-channel BO and spin-channel BO order parameters are given
by $\sum_{\tau\tau^\prime} \bigl< a_{i\tau}^\dagger a_{j\tau^\prime
}\bigr> $ and $\sum_{\tau\tau^\prime} \sigma_{\nu}^{\tau\tau^\prime}
\bigl< a_{i\tau}^\dagger a_{j\tau^\prime }\bigr> $ for nearest-neighbor
$\left<ij\right>$, respectively. \cite{comBO} Then, the instability of the
FL against, at least, magnetic, SC, and BO states should be examined in
this paper.

When $I_s(i\omega_l,{\bf q})$ is strong, the FL is unstable against a
magnetic state. The N\'{e}el temperature $T_N$ is defined as the highest
value of $T_N$ determined by 
$[\chi_s(0,{\bf q})]_{T=T_N} \rightarrow +\infty$ as a function of ${\bf
q}$, with $\chi_s(0,{\bf q})$ given by Eq.~(\ref{EqKondoKai}). When
$I_s(i\omega_l,{\bf q})$ is so weak that 
$[\chi_s(0,{\bf q})]_{T=0\hspace{1pt}{\rm K}} < +\infty$ for any ${\bf
q}$, the FL is stable against any magnetic state.

When $I_s^*(i\omega_l,{\bf q})$ is weak or strong, the FL is unstable
against an anisotropic superconducting (SC) state at least at $T=0$~K,
provided that no disorder exists. When $n\simeq 1$ or $U/W$ is not so
large, $I_s^*(i\omega_l,{\bf q})$ is antiferromagnetic. In such a case,
the FL is unstable against a singlet SC state. It is possible that
$I_s^*(i\omega_l,{\bf q})$ is ferromagnetic if the superexchange
interaction is very weak and the chemical potential is at the top or
bottom of the quasi-particle band, that is, if $U/W \gg 1$ and $n\simeq 0$
or $n\simeq 2$. In this case, the FL is unstable against a triplet SC
state.

The FL can also be unstable against a BO state and a flux state, which is
simply a multi-{\bf Q} BO state with different phases for different {\bf
Q} components. Within Kondo-lattice theory, magnetic or SC states are more
stable than BO and flux states are.

The above analysis cannot exclude possibility of a more exotic state. If
the exotic state is characterized by an order parameter and the order
parameter is specified, it is straightforward to examine the instability
of the FL against the exotic state by Kondo-lattice theory.

When $U/W\alt 1$, the conventional perturbation in terms of $U$ is more 
useful than that in terms of $I_s(i\omega_l,{\bf q})$. When the nesting
of FS is sharp, a non-interacting electron gas is unstable gainst a spin
density wave. When an interaction between electrons given by
$U^2\chi_s(i\omega_l,{\bf q})$ is considered, the electron gas is unstable
against an anisotropic SC state at least at $T=0~$K, provided that no
disorder exists.

No order parameter appears in one dimension. However, the FL that is
constructed under SSA and is stabilized beyond SSA can be used as an {\it
unperturbed} state to study one dimension by Kondo-lattice theory. The FL
for $U/W\agt1$ becomes a Tomonaga-Luttinger liquid except for $n=1$ and
$W/U=+0$ when $I_s(i\omega_l,{\bf q})$ is perturbatively treated, as the
electron gas does when $U$ is perturbatively treated. It is plausible that
Lieb and Wu's insulating state \cite{Lieb-Wu} for $n =1$ and $U \ne 0$ can
only be obtained by non-perturbative theory; the point of $U = 0$ is an
essential singularity.\cite{Takahashi} 



\section{Relevant effects for actual metal-insulator transitions}
\label{SecBeyondHub}

Since no M-I transition occurs at finite $U$ in two dimensions and
higher, actual M-I transitions cannot be explained within the Hubbard
model. Therefore, various effects should be considered in a multi-band or
multi-orbital model. Changes of lattice symmetries or jumps in lattice
constants are often observed, \cite{tokura} which implies that the
electron-phonon interaction should also be considered in the multi-orbital
model. It is likely that a relevant electron-phonon interaction arises from
spin channels \cite{el-ph1,el-ph2} and orbital channels rather than charge
channels because local charge fluctuations are suppressed, as is discussed
in Sec.~\ref{SecFL-Relation}. Cooperative Jahn-Teller or orbital ordering
must be responsible for the change of lattice symmetries. Not only the
electron-phonon interaction but also the orbital-channel exchange
interaction \cite{inagaki,cyrot,itinerant-ferro} can play a role in the
orbital ordering, as a spin-channel exchange interaction is responsible
for a spin or magnetic ordering.

The FS sum rule holds for the quasi-particles; the ordinary rule holds in
the absence of an order parameter, and a modified rule holds even when
the Brillouin zone is folded by an antiferromagnetic or orbital order
parameter. Since a crystalline solid is a metal provided the Fermi surface
is present while it is an insulator provided that the Fermi surface is
absent, Wilson's classification of crystalline solids into metals and
insulators \cite{wilson} applies to M-I transitions. Two types of M-I
transitions are possible according to the band structures of the
quasi-particles in the absence and presence of an order parameter: 
between a metal and an insulator and between a compensated metal and an
insulator.

The Kondo temperatures $T_K$ or $k_BT_K$ corresponds to the effective
Fermi energy of the quasi-particles. The Kondo temperatures $T_K$ can be
different in metallic and insulating phases of a first-order M-I
transition, provided that symmetries of the lattice or lattice constants
are changed. In the metallic phase, $T_K$ is higher than $T$ and the
quasi-particles are well defined. In the insulating phase, $T_K$ is lower
than $T$ so that the quasi-particles are not well defined. In such a case,
the M-I transition is a transition between a high-$T_K$ itinerant-electron
phase and a low-$T_K$ local-moment phase. Change of lattice symmetries or
jumps in lattice constants must play a crucial role in any first-order M-I
transition, in particular, in a metal-insulator transition between the
high-$T_K$ phase and the low-$T_K$ phase.

Since disorder, either small or large, must always exist, Anderson
localization can play a role in M-I transitions or crossovers. The
broadening of the quasi-particle band, which is examined in
Sec.~\ref{SecBroadening}, depends on disorder. \cite{phase-diagram} The
band broadening in the presence of disorder can also play a role in actual
M-I transitions or crossovers. 

\section{Magnetism crossover}
\label{SecMagnetismCrossover}
The N\'{e}el temperature $T_N$ can be nonzero in three dimensions and
higher. Even in one and two dimensions, there exists a temperature scale
$T_N^*$, below which critical thermal fluctuations are developed;
$T_N^* \simeq T_N$ in three dimensions and higher. In accordance with the
$T$-dependent crossover between a localized spin for $T\gg T_K$ and a
normal FL for $T\ll T_K$ in the Kondo problem, \cite{wilsonKG} magnetism
for $T_N^* \gg T_K$ is characterized as typical local-moment magnetism 
and magnetism for $T_N^* \ll T_K$ is characterized as typical
itinerant-electron magnetism. \cite{phase-diagram} The magnetism crossover
is simply a Mott M-I crossover between an insulating magnet at $T\agt T_K$
and a metallic magnet at
$T\alt T_K$.
 
According to Eq.~(\ref{EqKondoKai}), possible mechanisms for the
Curie-Weiss (CW) law are the temperature dependences of
$\tilde{\chi}_s(0)$, $J_Q(0,{\bf q})$, and $ -4 \Lambda (0,{\bf q})$; the
temperature dependence of the superexchange interaction
$J_s({\bf q})$ can be ignored at $T\ll U/k_B$. No other mechanism is
possible.

In local-moment magnets at $T\agt T_K$, the quasi-particles are not well
defined so that $J_Q(0,{\bf q})$ is vanishing. \cite{three-exchange} The
local susceptibility $\tilde{\chi}_s(0)$, which is nonzero even in infinite
dimensions, shows the CW law for any ${\bf q}$, which is characteristic of
the CW law of local-moment magnets. The mode-mode coupling term $- 4
\Lambda (0,{\bf q})$, which vanishes in infinite dimensions, can modify
the CW law in finite dimensions.

In itinerant-electron magnets at $T\alt T_K$, the quasi-particles are well
defined so that $J_Q(0,{\bf q})$, which can be nonzero for particular
${\bf q}$ corresponding to magnetic Weiss mean fields even in infinite
dimensions, is responsible for the CW law.\cite{miyai} When there is a
sharp nesting of the Fermi surface, $J_Q(0,{\bf q})$ shows a temperature
dependence consistent with the CW law for only ${\bf q}$'s close to the
nesting wave vector. When the chemical potential lies around a sharp peak
of the density of states, $J_Q(0,{\bf q})$ shows a temperature dependence 
consistent with the CW for only small ${|\bf q}|\simeq 0$. Such ${\bf q}$
dependences are characteristic of the CW law of itinerant-electron
magnets. On the other hand, the mode-mode coupling term $ -4 \Lambda
(0,{\bf q})$ gives an inverse CW temperature dependence or it suppresses
the CW law in finite dimensions. \cite{miyai,miyake}

\section{High-$T_c$ superconductivity}
\label{SecSuperconductivity}
According to the resonating-valence-bond (RVB) theory of high-$T_c$
superconductivity, \cite{RVB} the {\em normal} state above $T_c$ is the
RVB state in cuprate superconductors, which lie in the vicinity of the
Mott M-I transition. The RVB state is stabilized by the formation of an
itinerant or resonating singlet on each pair of nearest-neighbor unit
cells due to the superexchange interaction. On the other hand, it is shown
in Sec.~\ref{SecBroadening} of this paper that the FL is stabilized by
the Fock-type term of the superexchange interaction or, physically, by
the formation of an itinerant singlet on each pair of nearest-neighbor
unit cells. The stabilization mechanisms are, at least, similar to each
other in the RVB theory and Kondo-lattice theory. 

If the RVB state is characterized by an order parameter and the order
parameter is specified, it is straightforward to examine the instability
of the FL against the RVB state by Kondo-lattice theory. However, no order
parameter has been proposed so far, at least, within a {\em real-electron}
model, i.e., the Hubbard or $t$-$J$ model. \cite{comHolon} It is proposed
therefore in this paper that the symmetry of the RVB state is not broken
and is the same as that of the normal FL. On the basis of adiabatic
continuity,\cite{AndersonText} the stabilized FL is simply an RVB state 
provided that it is mainly stabilized by the RVB effect or 
$c_J|J| \gg W/\tilde{\phi}_\gamma$ in Eq.~(\ref{EqW*}).
 According to Kondo-lattice theory, the cooperation between the Kondo
effect, which favors a local singlet on each unit cell, and the RVB
effect, which favors a local singlet on each pair of nearest-neighbor
unit cells, stabilizes the normal Fermi liquid, whose ground state is a
singlet as a whole. The stabilized normal FL is simply the normal state
above $T_c$ of cuprate superconductors. 

Experimentally, the superexchange interaction constant of cuprate 
superconductors is as large as $J=-(0.10\mbox{--}0.15)~\mbox{eV}$ between
nearest neighbors. When nonzero bandwidths of LHB and UHB are considered,
it follows that $|J |\simeq 0.5 \times 4|t|^2/U$, as is discussed in
Sec.~\ref{SecKondo1/D}. Since $|J |\alt0.08\mbox{~eV}$ for actual
$t\simeq-0.4~\mbox{eV}$ and $U\agt 4~\mbox{eV}$, it is difficult to
reproduce consistently such $J$ within the Hubbard model.
\cite{exchange-reduction} Then, the $d$-$p$ model or the $t$-$J$ model
should be used instead of the Hubbard model in order to explain high-$T_c$
superconductivity quantitatively. \cite{exchange-reduction} It is
straightforward to develop Kondo-lattice theory for the $d$-$p$ model and
the $t$-$J$ model.

According to an early FL theory of high-$T_c$ superconductivity, 
\cite{highTc1,highTc2} the condensation of $d\gamma$-wave Cooper pairs of
the Gutzwiller quasi-particles due to the superexchange interaction is
responsible for high-$T_c$ superconductivity. It is analyzed in this paper
that Kondo-lattice theory is simply FL theory, in which a normal FL is an
{\it unperturbed} state within the Hilbert subspace with no order 
parameter and a true ground state is studied in the whole Hilbert space with no
restriction. The analysis confirms the early theory. 

The analysis also confirms theories of anomalous or exotic properties of 
cuprate oxide superconductors, which treat the softening of phonons
caused by antiferromagnetic spin fluctuations, 
\cite{el-ph1,el-ph2} $4a$-period stripes or $4a\times 4a$-period checker
boards caused by $8a$-period or $8a\times 8a$-period spin density wave
(SDW), \cite{mag-structure,el-ph2} the opening of pseudogaps above $T_c$,
\cite{psgap1,psgap2} nonzero-${\bf Q}$ or multi-${\bf Q}$
superconductivity in the presence of the stripe or checker-board order,
\cite{ztpg} with ${\bf Q}$ being the total momenta of Cooper pairs here,
and the suppression of the specific heat coefficient $\gamma$ in the
region of the Mott M-I crossover, which is examined in
Appendix~\ref{SecAppendixRho} of this paper.


\section{Discussion}
\label{SecDiscussion}

The occurrence of a first-order M-I transition at $T>0$~K is suggested by
a numerical SSA theory or DMFT not only for $n=1$ but also for $n\ne 1$.
\cite{RevMod,PhyToday,kotliar} A similar phenomenon to that observed by
the numerical DMFT is also observed at $T>0$~K by a Monte Carlo theory,
\cite{imada} which is beyond SSA. In these numerical theories, the static
homogeneous charge susceptibility or the compressibility 
\begin{equation}
\chi_c(i\omega_l=0,|{\bf q}|\rightarrow 0) = dn(\mu)/d\mu, 
\end{equation}
shows a rapid change. When the rapid change is really a jump, the phase
diagram for $T>0$~K is like that shown in Fig.~1 of 
Ref.~\onlinecite{kotliar}. The phase diagram suggests that the first-order
M-I transition occurs even at $T=0$~K. However, the first-order M-I
transition at $T=0$~K is inconsistent with the second-order transition
within SSA predicted by Brinkman and Rice's theory \cite{brinkman} and
the analysis of this paper. It is interesting to clarify the nature of the
rapid jump observed by the numerical theories, whether it is really a
transition or a sharp crossover between $T\gg T_K$ and $T\ll T_K$. If the
rapid jump is really a first-order transition, it is interesting to
examine whether or not, as temperatures go down to $T=0$~K, the
first-order M-I transition turns over to a first-order metal-metal
transition, which is discussed in Sec.~\ref{SecMMtransition}.

When $U/W \agt 1$, charge fluctuations are suppressed within SSA, as is
discussed in Sec.~\ref{SecFL-Relation}. Since the {\it unperturbed} state
of Kondo-lattice theory is the normal FL constructed in SSA, it is
unlikely that the divergence of the charge susceptibility occurs. Within
Kondo-lattice theory, it is difficult for the FL to be unstable against 
the gas-liquid type M-I transition, at least, driven by the divergence of
charge-density fluctuations. \cite{misawa}

The long range Coulomb interaction exists in actual solids. Since it
requires the charge neutrality, the electron density $n$ must be kept
constant so that the compressibility identically vanishes such that
$dn(\mu)/d\mu=0$. The compressibility can never be any relevant property
for actual M-I transitions. 

\section{Conclusion}
\label{SecConclusion}
The Hubbard model in the strong-coupling regime is mainly studied by 
Kondo-lattice or $1/d$ expansion theory, with $d$ being the spatial dimensionality.
Relevant leading-order effects in $1/d$  are local spin fluctuations 
and magnetic Weiss mean fields. Local spin
fluctuations are considered in the best single-site approximation (SSA), which
is reduced to a problem of determining and solving self-consistently the
Anderson model and is rigorous for $d\rightarrow +\infty$ but within the
Hilbert subspace with no order parameter. 
Multi-site or intersite effects, which include not only 
magnetic Weiss mean fields but also higher-order effects in $1/d$, are
perturbatively considered beyond SSA.

In two dimensions and higher, the ground state within the Hilbert subspace 
with no order parameter is a normal Fermi liquid except for $n=1$ and 
$W/U=+0$, with $n$ being the electron density per unit cell, $W$ 
the bandwidth, and $U$ the on-site repulsion. In the strong coupling regime of
$U/W\agt 1$, the Fermi-liquid  ground state is stabilized 
by the cooperation  between the Kondo effect and 
the resonating-valence-bond effect, i.e., 
the quenching of magnetic moments by {\em single-site} and {\em
nearest-neighbor} local quantum spin fluctuations. 
In the whole Hilbert space with no restriction, eventually, the normal
Fermi liquid is unstable at least against a magnetic or superconducting
state except for a trivial case of $U=0$. On the other hand, the ground
state for $n=1$ and $W/U=+0$ is a typical Mott insulator.
 
In one dimension, the ground state is a Tomonaga-Luttinger liquid except 
for $n=1$ and $W/U=+0$. Lieb and Wu's insulating state cannot be reproduced
by the perturbative treatment of intersite effects in this paper.

Since actual metal-insulator transitions cannot be explained within the
Hubbard model, in order to explain them, one or several effects among 
the electron-phonon interaction, multi-band or multi-orbital effects, and
effects of disorder should be considered beyond the Hubbard model. In
particular, change of lattice symmetries or jumps in lattice constants
must play a crucial role in any first-order metal-insulator transition.

The energy scale of local quantum spin fluctuations is the Kondo
temperature $T_K$ or $k_BT_K$. The Gutzwiller quasi-particles are well
defined in the high-$T_K$ phase, which is defined by $T \alt T_K$.
Whether a crystalline solid in the high-$T_K$ phase is a metal or an
insulator can be explained by the extended Wilson's classification of the
band structure of the quasi-particles in the absence or presence of an
order parameter; the solid is a metal provided the Fermi surface is
present while it is an insulator provided that the Fermi surface is
absent. On the other hand, a crystalline solid in the low-$T_K$ phase, 
which is defined by $T \agt T_K$, is an insulator.

The crossover between local-moment magnetism and itinerant-electron
magnetism is simply a Mott metal-insulator crossover between a metallic
magnet at $T\alt T_K$ and an insulating magnet at $T\agt T_K$. Typical
local-moment magnetism and itinerant-electron magnetism are therefore 
characterized by $T_N^* \gg T_K$ and $T_K \gg T_N^*$, respectively, with
$T_N^*$ being a temperature scale of magnetism, below which magnetic order
parameter appears or critical spin fluctuations are well developed.

In fact, Kondo-lattice theory is a Fermi-liquid theory, in which a normal
Fermi liquid is constructed as an {\it unperturbed} state within the
Hilbert subspace with no order parameter and a true ground state is
studied in the whole Hilbert space with no restriction. The analysis by
Kondo-lattice theory confirms the early Fermi-liquid theory
\cite{highTc1,highTc2} of high-temperature superconductivity.

\begin{acknowledgments}
The author is thankful to M. Ido, M. Oda, and N. Momono for useful
discussions on the specific heat coefficient and the asymmetry of
tunneling spectra of cuprate oxide superconductors.
\end{acknowledgments}


\appendix
\section{Proof of the inequality} 
\label{SecProof}
When the following real functions,
\begin{equation}
S_1(\varepsilon,\varepsilon^\prime) =
\varepsilon +\mu - \epsilon_a - \varepsilon^\prime 
- \mbox{Re} \left[\tilde{\Sigma}_\sigma(\varepsilon+i0)\right],
\end{equation}
\begin{equation}
S_2(\varepsilon)= \delta \Delta 
- \mbox{Im} \left[\tilde{\Sigma}_\sigma(\varepsilon+i0)\right] ,
\end{equation}
and 
\begin{equation}
Y_n(\varepsilon) = \int d\varepsilon^\prime
\rho_0(\varepsilon^\prime) 
\frac{S_1^n(\varepsilon,\varepsilon^\prime) }
{S_1^2(\varepsilon,\varepsilon^\prime) +S_2^2(\varepsilon)},
\end{equation}
with $n$ being an integer here, are defined, the single-site Green function
(\ref{EqR-delta}) is given by
\begin{eqnarray}
R_\sigma(\varepsilon+i0) &=&
%
Y_1(\varepsilon) - i S_2(\varepsilon) Y_0(\varepsilon) .
\end{eqnarray}
It follows from the mapping condition (\ref{EqMappingCondition}) that
\begin{eqnarray}
\mbox{Im}L_\sigma(\varepsilon+i0) &=& 
%
- \delta \Delta + \frac{S_2(\varepsilon)}
{Y_1^2(\varepsilon) + S_2^2(\varepsilon) Y_0^2(\varepsilon)} 
\nonumber \\ && \times
\left[Y_1^2(\varepsilon) - Y_0(\varepsilon) Y_2(\varepsilon)
 \right], \qquad 
\end{eqnarray}
where a relation of 
\begin{equation}
Y_2(\varepsilon) =1 - S_2^2(\varepsilon) Y_0(\varepsilon) ,
\end{equation}
is made use of. It should be noted that
$Y_0(\varepsilon)$, $Y_1(\varepsilon)$, $Y_2(\varepsilon)$, and
$L_\sigma(\varepsilon+i0) $ are continuous and finite provided that
$\delta>0$, even if $\rho_0(\varepsilon)$ or
$\tilde{\Sigma}_\sigma(\varepsilon+i0)$ is discontinuous or divergent.

Since the following inequality
\begin{equation}\label{EqInequality1}
\int d\varepsilon^\prime \rho_0(\varepsilon^\prime)
\frac{ \left[x+ S_1(\varepsilon,\varepsilon^\prime) \right]^2 }{
S_1^2(\varepsilon,\varepsilon^\prime) +S_2^2(\varepsilon)} >0,
\end{equation}
or 
\begin{equation}\label{EqInequality2}
Y_0(\varepsilon) x^2 + 2 Y_1(\varepsilon) x + Y_2(\varepsilon) >0,
\end{equation}
is satisfied for any real $x$, the discriminant should be negative in such a way that
\begin{equation}
Y_1^2(\varepsilon)-Y_0(\varepsilon)Y_2(\varepsilon) < 0.
\end{equation}
It is obvious that
\begin{equation}
S_2(\varepsilon) \ge \delta \Delta ,
\end{equation}
and
\begin{equation}
Y_1^2(\varepsilon) + S_2^2(\varepsilon) Y_0^2(\varepsilon) \ge 0.
\end{equation}
Therefore, it follows that
\begin{equation}\label{EqInequality3}
\mbox{Im}L_\sigma(\varepsilon+i0) \le - \delta \Delta ,
\end{equation}
even if $\tilde{\Sigma}_\sigma(\varepsilon+i0)$ is discontinuous or divergent.
The inequality~(\ref{EqImp}) is proved. 

 
\section{Quasi-particle in cuprate oxide superconductors}
\label{SecAppendixRho}

The density of states (DOS) $\rho^*(0)$ of the quasi-particle band is not
renormalized within SSA but is renormalized beyond SSA, as is examined in
Secs.~\ref{SecFL-Relation} and \ref{SecBroadening}. This issue is
phenomenologically considered in this Appendix.

The self-energy is expanded as
\begin{equation}
\Sigma_{\sigma}(\varepsilon+i0, {\bf k}) =
\Sigma_0({\bf k}) + \left[1 - \phi_\gamma({\bf k})\right] \varepsilon 
+ O\left(\varepsilon^2\right),
\end{equation}
at $T=0$~K.
Then, the electron density $n$ is given by 
\begin{equation}\label{EqAppFS-sum}
n = \frac1{N} \sum_{{\bf k}\sigma}
\theta\bigl( [ \mu - E({\bf k}) - \Sigma_0({\bf k}) ]/W \bigr) ,
\end{equation}
DOS at the chemical potential is given by
\begin{equation}
\rho^*(0) 
%
= \frac1{N} \sum_{{\bf k}}
\delta\left[\mu - E({\bf k}) - \Sigma_0({\bf k}) \right] ,
\end{equation}
and the specific heat coefficient is given by
\begin{equation}
\gamma =\frac{2}{3} \pi^2 k_B^2 
\frac1{N} \sum_{{\bf k}}\phi_\gamma ({\bf k})
\delta\left[\mu - E({\bf k}) - \Sigma_0({\bf k}) \right] .
\end{equation}
%
According to Eq.~(\ref{EqAppFS-sum}), 
any physical property can be regarded as a function of
$n$ instead of $\mu$.

Provided that the ${\bf k}$ dependence of $\Sigma_0({\bf k}) $
can be ignored, $\rho^*(0)$ does not depend on $U$.
In this case, 
\begin{subequations}\label{EqNonRenor}
\begin{equation}\label{EqRoh*AppB}
\rho^*(0) \simeq 1/W, 
\end{equation}
and
\begin{equation}\label{EqGam*AppB}
\gamma \simeq
\frac{2}{3} \pi^2 k_B^2 \frac{1} {W^*}, 
\end{equation}
with 
\begin{equation}\label{EqW*App}
W^* = W/
\left<\phi_\gamma ({\bf k})\right>_{\rm FS} ,
\end{equation}
\end{subequations}
being an effective bandwidth of the quasi-particles. Here,
$\left<\phi_\gamma ({\bf k})\right>_{\rm FS}$ is an average over FS. The
$n$ dependence of $\gamma$ mainly arise from that of $\left<\phi_\gamma
({\bf k})\right>_{\rm FS}$ in this case; it is certain that
$\left<\phi_\gamma ({\bf k})\right>_{\rm FS}$ increase as $n$ approaches
unity.

According to observations, \cite{loram,momono} 
$\gamma$ of hole-doped cuprate superconductors shows a peak around
$n\simeq 0.80\mbox{-}0.85$ with a peak value about $\gamma
\simeq 15$~mJ/mol~K$^2$. It decreases as $n$ approaches unity; $\gamma
\simeq 5$~mJ/mol~K$^2$ for $0.95\agt n \agt 0.90$. It is difficult to
explain such an $n$ dependence of $\gamma$ in terms of the $n$ dependence
of $\left<\phi_\gamma ({\bf k})\right>_{\rm FS}$. Only the possible
explanation by FL theory for the suppression of $\gamma$ for $n\simeq 1$
is the band broadening caused by the dispersion of 
$\Sigma_0({\bf k})$. In this appendix, a broadening factor $c_W(n)$ is
phenomenologically introduced such that
\begin{subequations}\label{EqRenor}
\begin{equation}\label{EqAppB1}
\rho^*(0;n) = 1/\left[c_W(n) W\right] , 
\end{equation}
and 
\begin{equation}\label{EqGam*AppB2}
\gamma (n) \simeq
\frac{2}{3} \pi^2 k_B^2 \frac1{W^*(n)},
\end{equation}
with
\begin{equation}\label{EqAppB2}
W^*(n) = c_W(n) W/
\left<\phi_\gamma ({\bf k};n)\right>_{\rm FS} ,
\end{equation}
\end{subequations}
where the $n$ dependences are explicitly shown.

For the sake of simplicity, two typical cases of hole dopings such as
$1-n=0.08$ and $1-n=2 \times 0.08$, that is,
$n=0.92$ and $n=0.84$ are considered; $n\simeq 0.84$ is an optimal doping,
where superconducting critical temperatures $T_c$ show a maximum as a
function of $n$. According to the Gutzwiller approximation, it follows that
$\left<\phi_\gamma ({\bf k};n\simeq 0.92)\right>_{\rm FS}/
\left<\phi_\gamma ({\bf k};n\simeq 0.84)\right>_{\rm FS} \simeq 2$. 
According to observations, \cite{loram,momono}
$\gamma(n\simeq 0.92)/\gamma(n\simeq 0.84) \simeq 1/3$, as is discussed
above. Then, it follows that
\begin{equation}
\frac{\rho^*(0;n\simeq 0.92)} 
{\rho^*(0;(n\simeq 0.84)} =
\frac{c_W(n\simeq 0.84)}{c_W(\simeq 0.92)}
 \simeq \frac1{6}. 
\end{equation}
The observed $n$ dependence of $\gamma$ implies that $\rho^*(0;n)$ must be
significantly suppressed for $n\simeq 1$. 

In hole-doped cuprates with $1>n\agt 0.8$, the chemical potential lies
around the band center of the Gutzwiller band, just below which LHB is
present. According to Eq.~(\ref{EqGam*AppB}) or (\ref{EqGam*AppB2}), a
half of the Gutzwiller bandwidth is
$W^*/2 \simeq 0.12\mbox{-}0.15~\mbox{eV}$ for $\gamma \simeq
15$~mJ/mol~K$^2$ and is $W^*/2 \simeq 0.35\mbox{-}0.50~\mbox{eV}$ for
$\gamma \simeq 5$~mJ/mol~K$^2$. Since it is rather small, not only the
Gutzwiller band but also the top part of LHB can be observed by tunneling
spectroscopy. According to the Hubbard approximation, DOS at the center of
LHB is given by
\begin{equation}\label{EqRhoLHB} 
-(1/\pi)\mbox{Im} R_\sigma(\epsilon_a - \mu +i0) 
\simeq O \left(1/W \right),
\end{equation} 
and DOS at an energy sightly deeper than the top of LHB is still $O(1/W)$. 
It is certain that DOS of LHB never drastically changes by a slight
change of dopings. On the other hand, the suppression of $\rho^*(0)$
occurs as $n$ approaches unity; $\rho^*(0)$ given by
Eq.~(\ref{EqRoh*AppB}), which is not suppressed, is as large as DOS of
LHB given by Eq.~(\ref{EqRhoLHB}). The different $n$ dependences of DOS
between the Gutzwiller band and LHB must be responsible for the observed
asymmetry of tunneling spectra. 
\cite{asymmetry1}

The contribution to $\Delta\Sigma_0({\bf k})$ by the Fock-type term of the
superexchange interaction is only considered in Sec.~\ref{SecBroadening}. 
It is interesting to examine microscopically how large contribution to 
$\Delta\Sigma_0({\bf k})$ can arise from three types of fluctuations such
as antiferromagnetic, superconducting, and charge-channel BO fluctuations,
which must compete with each other, as is discussed in
Sec.~\ref{SecInstabilityFL}.


\begin{thebibliography}{}
\bibitem{com1/D}
In any single-band model such as the Hubbard model, leading-order effects
in $1/d$ relevant in repulsive systems ($U/W>0$) are local spin
fluctuations and Weiss mean fields of magnetism. Leading-order effects in
$1/d$ relevant in attractive systems ($U/W< 0$) are isotropic $s$-wave or
BCS local superconducting fluctuations, local charge fluctuations, and
Weiss mean fields of BCS superconductivity and charge density wave (CDW).
All the other effects are of higher order in $1/d$. Since Weiss mean
fields are intersite effects, the appearance of an order parameter cannot
be treated in SSA but can be treated by Kondo-lattice or $1/d$ expansion
theory. It is quite reasonable that no broken symmetry occurs under SSA,
as is examined in Sec.~\ref{SecSSA}.
\bibitem{comHolon}
According to a slave-boson RVB theory, the RVB state above $T_c$ is
characterized by a certain order parameter for auxiliary particles, which
are called spinons and holons, in the auxiliary-particle or slave-boson
$t$-$J$ model. However, it can never be shown or, at least, it has never
been shown so far what symmetry is actually broken above $T_c$ in the
corresponding $t$-$J$ model when the order parameter appears in the
slave-boson $t$-$J$ model.
\end{thebibliography}
\end{document}